# Deep residual network for sound source localization in the time domain

*Dmitry Suvorov* [1], *Ge Dong* [2] and *Roman Zhukov*[1]

1. Center for space research, Skolkovo Institute of Science and Technology, Moscow 143026, Russia;

2. School of Aerospace Engineering, Tsinghua University, Beijing 100084, China

**Abstract:** This paper presents a system for sound source localization in time domain using a deep residual neural network. Data from the linear 8-channel microphone array with 3 cm spacing is used by the network for direction estimation. We propose to use the deep residual network for sound source localization considering the localization task as a classification task. This article describes the gathered dataset and developed architecture of the neural network. We will show the training process and its result in this article. The developed system was tested on validation part of the dataset and on new data capture in real time. The accuracy classification of 30 ms sound frames is 99.2%. The standard deviation of sound source localization is 4 degrees. The proposed method of sound source localization was tested inside of speech recognition pipeline. Its usage decreased word error rate by 1.14% in comparison with similar speech recognition pipeline using GCC-PHAT sound source localization.

## INTRODUCTION

The purpose of the work is to develop a sound source localization system based on data obtained from a linear compact microphone array. The system should be resistant to noise and reverberation and also should be able to work in real time on conventional personal computers.

A large amount of noise and reverberation in captured sound is the key problem for distant speech recognition systems [1]. To solve this problem, a sound signal can be captured by microphone array to perform sound source localization and beamforming. In this case the full process of sound capture and processing will consist of the following steps [2]:

- Sound capture with microphone array.
- Sound source localization and tracking.
- Beamforming.
- Post-filtering.

Sound source localization is the key element in this architecture because its accuracy defines quality of algorithms for implementation at further stages. Beamforming and post-filtering use previously defined sound source direction as input parameter.

At the moment there are a large number of methods for sound source localization:

- Weighted GCC-PHAT [3] and its analogs, which use sound channels correlation. The baseline version of GCC-PHAT is presented in Eq. 1 and Eq. 2.

$$GCC_{kl}(\tau) = \int \frac{Y_k(\omega)Y_l(\omega)e^{j\omega}}{|Y_k(\omega)||Y_l(\omega)|} d\omega \quad (1)$$

, where $Y_k(\omega)$ and $Y_l(\omega)$ are discrete Fourier transforms of k and l channels of the sound frame from the microphone array.

Likelihood of presence of active sound source at direction $\Theta_i$:

$$loglik(\theta_i) = \frac{1}{M} \sum_{kl}^{M} GCC_{kl}(\tau_{kl}^*(\theta_i)) \quad (2)$$

, where M is a number of channels, $\Theta$ is a direction (azimuth for a linear microphone array, azimuth and elevation for planar and 3D configurations), $\tau_{kl}^*(\Theta_i)$ is a theoretical delay between k and l channels for $\Theta_i$ direction of arrival.

- IDOA algorithms [4] estimating phase delays on different frequencies between channels of captured multichannel sound (Eq. 3 - 7).

Likelihood of presence of active sound source with frequency $\omega$ at direction $\Theta_i$:

$$loglik(\theta_i|\omega) = \frac{-||mod(\delta(\omega) - \Delta(\omega,\theta_i), 2\pi)||_2^2}{||\frac{\partial \Delta}{\partial \theta}(\omega,\theta_i)||} \quad (3)$$

, where $\Delta(\omega, \Theta_i)$ is a vector of theoretical phase differences between k and zero microphones at frequency $\omega$ for the active sound source located at direction $\Theta_i$ and $\delta(\omega)$ is a vector of measured phase differences between k and zero microphones at frequency $\omega$:

$$\delta_k(\omega) = \angle Y_k(\omega) - \angle Y_0(\omega) \quad (4)$$
$$\delta(\omega) = [\delta_1(\omega), \delta_2(\omega), \dots, \delta_{M-1}(\omega)] \quad (5)$$

Probability of presence of active sound source with frequency $\omega$ at direction $\Theta_i$:

$$P(\theta_i|\omega) = \frac{\exp(\frac{loglik(\theta_i|\omega)}{\sigma})}{\sum_j \exp(\frac{loglik(\theta_j|\omega)}{\sigma})} \quad (6)$$

The most probable direction to the wideband sound source:

$$\hat{\theta} = \text{argmax}_\theta (\sum_\omega P(\theta_i|\omega)) \quad (7)$$

- Scanning of the surrounding area with Delay-and-sum beamformer [5] or other types of beamformers.

Likelihood of presence of active sound source with frequency $\omega$ at direction $\Theta_i$ when scanning is performed using Delay-and-sum beamformer:

$$loglik(\theta_i|\omega) = \frac{1}{M} \sum_{m=0}^{M-1} e^{j\omega\tau_m^*(\theta_i)} Y_m(\omega) \quad (8)$$

The most probable direction to sound source can also be calculated using Eq. 6 and 7.

- MUSIC algorithms [6] and their modifications.

Probability of presence of active sound source with frequency $\omega$ at direction $\Theta_i$:

$$loglik(\theta_i|\omega) = \frac{1}{\alpha(\omega,\theta_i)^H(I - U_s U_s^H)\alpha(\omega,\theta_i)} \quad (9)$$

, where $\alpha(\omega, \Theta_i)$ is the capturing matrix with size M by J` (J is a number of sound sources), $U_s$ is signal subspace eigenvectors matrix [7]. The most probable direction to sound source can also be calculated using Eq. 6 and 7.

- Sound source localization algorithms based on deep neural networks [8] using convolutional and residual layers.

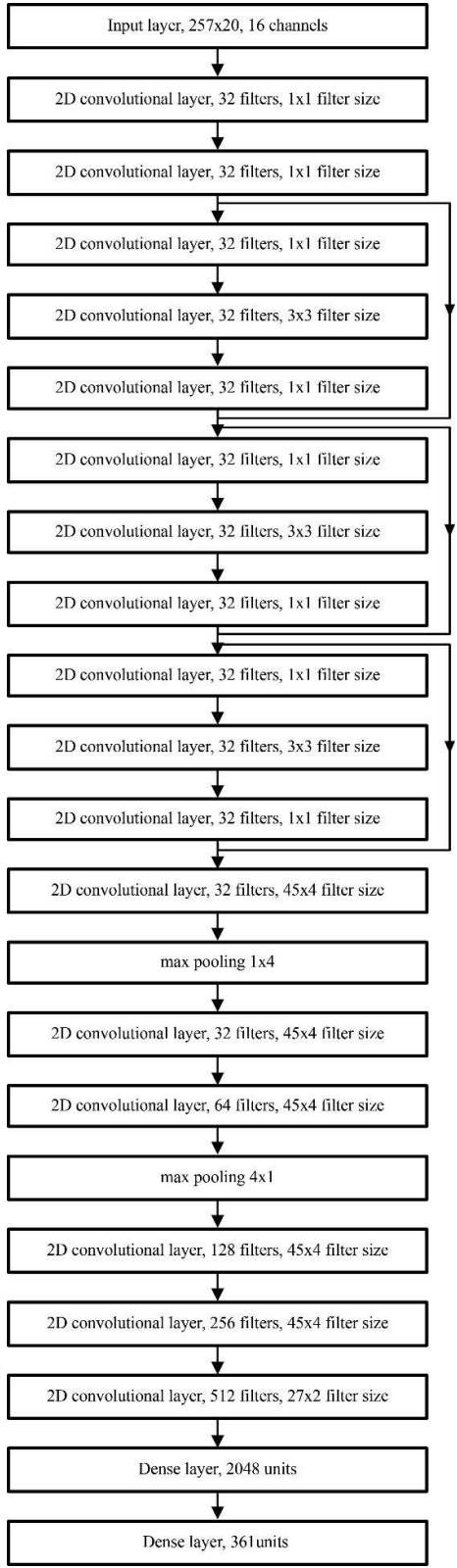

Fig. 1 Residual CNN proposed in [8]. Each convolutional layer is followed by the ReLU non-linearity.

Vector of probabilities of presence of sound source at possible directions:

$$P(\theta_0,\ldots,\theta_{N-1}) = F(Y_0(\omega),\ldots,Y_{M-1}(\omega)) \quad (10)$$

, where N is a number of checking directions. The architecture proposed by [8] is shown in Fig 1.

- Human speech localization algorithms based on processing data from microphone array and video camera [9].

All the methods except the one based on neural networks consider the localization problem as a problem of testing the hypothesis about sound source presence in a specific space sector, which leads to an increase in required computing power, as it is needed to check sound source presence in the surrounding space with a specified step [7]. Also, their implementations use assumptions about the plane front of an acoustic wave [7] which leads to errors in localization of sound sources located closely to a microphone array [10].

The method based on neural networks, described in [8], considers the localization problem as a problem of sound frame classification into sound source direction classes and the classification of active sound source absence. As input data, the algorithm uses the discrete Fourier transform (DFT) for every channel with sound duration of about several tens of milliseconds with some previous frames. Necessity in DFT computation for each channel on each iteration and use of two-dimensional convolutional layers, lead to increased computing complexity of the algorithm. Moreover, the algorithm uses only amplitude information from DFT and doesn't use phase information. It can also negatively affect the accuracy of localization.

Further in the paper, a sound source localization method based on deep convolutional neural networks using as input, multichannel sound frames with fixed duration from microphone array, will be proposed. Unlike in the method introduced in [8] the network uses only one-dimensional convolutions, which significantly reduces its computing complexity. Also the process of training dataset collection, neural network training and system testing will be described.

**MATERIALS AND METHODS**

**Dataset**: To perform experiments with deep neural network training, a big dataset of labeled data is required. To solve this problem a python application was developed, which plays sound via a speaker whilst simultaneously recording it with an 8-channel microphone array with 3 cm spacing, implemented on the basis of MEMS microphones with PDM interface [11] which is shown in Fig. 2. The application randomly chooses and plays a music file for a duration of 30 seconds from an array of one-channel sound files from "GTZAN Genre Collection" collected in the framework of [12]. In this way, one-hour multichannel sounds for each direction with a 10 degrees step from 0 to 180 degrees were recorded. One hour of silence was also recorded. Everything was recorded in a 2x3 m room. The sound was recorded with 16 kHz frequency with 16-bit resolution.

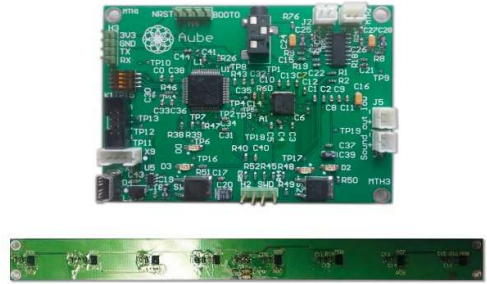

Fig. 2 Linear microphone array used for capturing the dataset and real time experiments with proposed sound source localization.

As dataset was collected with a linear microphone array, further the task of sound source localization was considered as a task for estimation of azimuth to sound source, because the use of linear microphone array makes it impossible to determine an elevation angle for obvious geometric reasons.

**Neural network architecture**: The developed neural network architecturally consists of four big blocks (Fig. 3):

- Input layer, accepting 8-channel sound frames from microphone array with duration of 480 samples (30 ms) in float format.
- First 1D convolutional layer [13], performing primary feature extraction (Eq. 11).
- Block consisting of two residual layers [14]. Residual layers allow a delay in overfitting of neural networks and therefore train deeper networks.

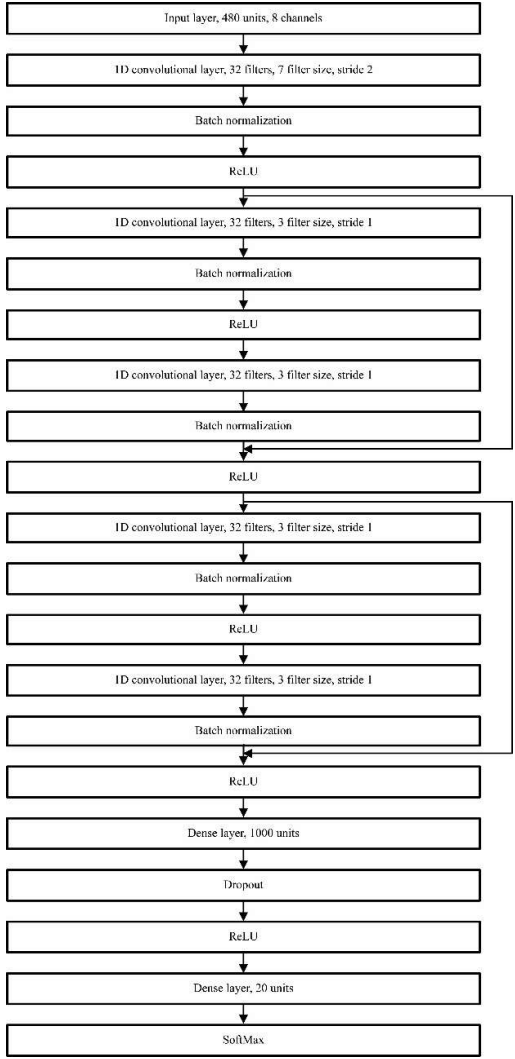

Fig. 3 A deep neural network, used for sound source localization.

- Decision-making blocks, consisting of two fully connected layers, create outputting probabilities that a sound frame has a sound from one of the possible azimuths and probability of absence of any active sound sources in the frame.

$$V(x,t) = \sum_{i=x-\frac{L-1}{2}}^{x+\frac{L-1}{2}} \sum_{s=1}^{S} K\left(i - x + \frac{L-1}{2}, s, t\right) U(i,s) \quad (11)$$

, where $U(x, s)$ is a 1D input signal containing S channels, t is number of output channel, $K(x, s, t)$ is a matrix of size L by S of the filter for t output channel.

After each convolutional layer, a batch norm layer is used to allow to train neural networks with a lesser number of iterations to postpone overfitting [15]. Batch Normalization Transform is shown in Eq. 12 - 15.

Mini-batch mean:

$$\mu_\beta = \frac{1}{m}\sum_{i=1}^{m} x_i \quad (12)$$

Mini-batch variance:

$$\sigma_\beta^2 = \frac{1}{m}\sum_{i=1}^{m}(x_i - \mu_\beta)^2 \quad (13)$$

Normalize:

$$\hat{x}_i = \frac{x_i - \mu_\beta}{\sqrt{\sigma_\beta^2 - \varepsilon}} \quad (14)$$

Final scale and shift:

$$y_i = \gamma\, x_i + \beta \quad (15)$$

, where m is a batch size, x is a batch of input data, $\gamma$ and $\beta$ are parameters to be learned.

After the first fully connected layer, a Dropout layer is also used to delay the moment of overfitting to later iterations [16]. Feed-forward operation of the dropout layer:

$$r_j^l = \text{Bernoulli}(p) \quad (16)$$

$$\hat{y}^l = r^l * y^l \quad (17)$$

, where $r^l$ is a vector of independent Bernoulli random variables each of which has probability p of being 1, * is an element-wise product.

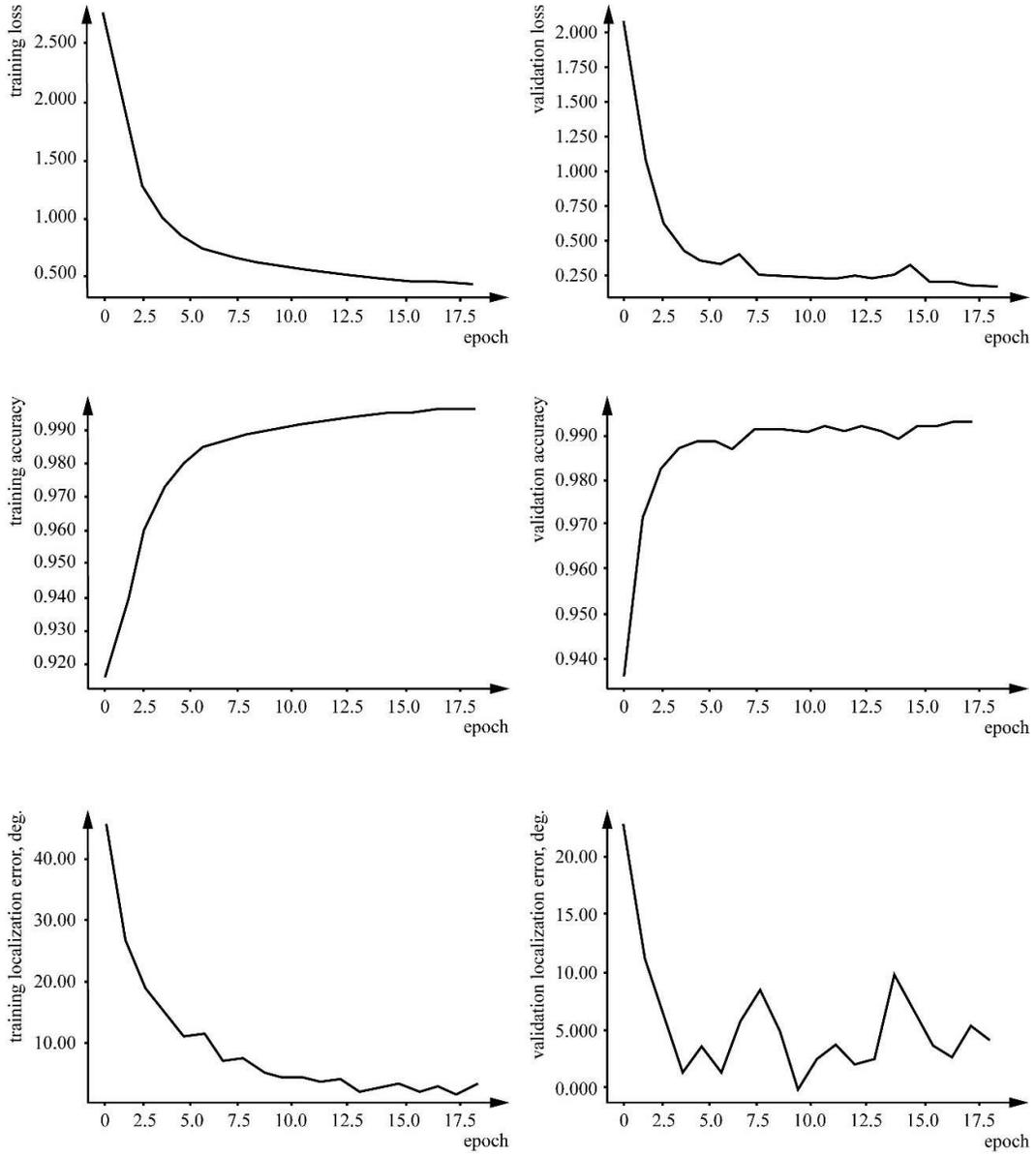

Fig. 4 The learning process of the developed system.

ReLU non-linearity [17] is used after all the convolutional and fully connecting layers with the exception of the last dense layer:

$$f(x) = \max(0, x) \qquad (18)$$

On the last layer SoftMax non-linearity [17], is used, as it is needed to normalize output of the neural network in such a way that the sum of all probabilities were equal to 1:

$$f(x_i) = \frac{e^{x_i}}{\sum_j e^{x_j}} \qquad (20)$$

Unlike the solution proposed in [8] the network accepts original signals but not its Fourier image. It is possible, as the Fourier transformation is essentially decomposition into narrowband components, and therefore one-dimensional convolutional layers are able to learn this decomposition themselves.

**Training and testing:** A prototype of the proposed system was realized with python based on the Theano and Lasagne libraries. The learning was done with the Adam optimization algorithm [19] with a low parameter of training speed (Eq. 21 - 23). The learning was done in 20 epochs on a NVIDIA GeForce GTX 1070 graphics card using Cuda technology. Cross-entropy was used as a loss function (Eq. 24). In the training process, the values of the loss function, the accuracy of the classification and the standard deviation of the azimuth determination error were monitored. In Fig. 4 it can be seen that overfitting happened only after 18 epochs.

$$w[t+1] = w[t] - \propto \frac{1}{\sqrt{g[t+1]+\varepsilon}} v[t+1] \quad (21)$$

$$g[t+1] = \mu\, g[t] + (1-\mu)\, \nabla(L, w[t])\, \nabla(L, w[t]) \quad (22)$$

$$v[t+1] = \beta\, v[t] + (1-\beta)\, \nabla(L, w[t]) \quad (23)$$

, where t is iteration number, L is loss function, w is set of trainable parameters of the network, ε, μ and β are scalar parameters of the algorithm.

$$L(p, y) = \frac{1}{N} \sum_{i=1}^{N} \sum_{j=1}^{N} y_{ij} \log(p_{ij}) \quad (24)$$

, where matrix p is NxM output of the neural network, matrix y is a one-hot encoded real class identifier, M is set to a number of classes, N is a batch size.

To analyze the training results t-SNE visualization [20] for features generated by the penultimate fully connected layer was implemented (Fig. 5). It clearly shows the cluster structure of features and the mutual arrangement of clusters corresponding to the spatial arrangement of real azimuths, which indicates the good quality of the neural network training.

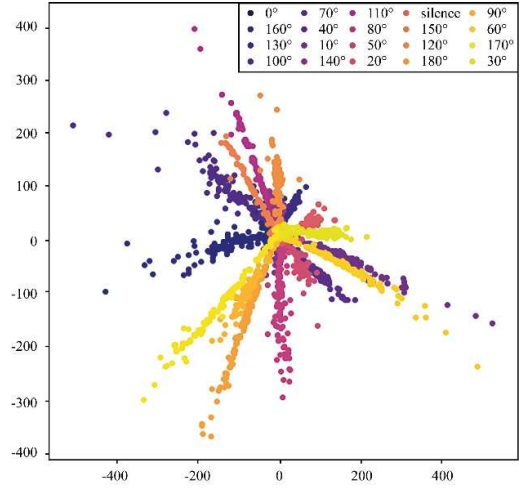

Fig. 5 t-SNE visualization of penultimate layer features.

## RESULTS AND DISCUSSION

**Evaluation in real time:** An application was developed which allows sound capture from the microphone array in real time and determines the direction of the sound source azimuth. Using this application, sound source directions were calculated in real time for sound sources in a previously known position. Measurements were done in a room where the training dataset was recorded and in another room that had a significantly different area and filling meaning it had different reverberation parameters. It can be seen in Fig. 6 that average absolute values of sound source direction azimuth determination error did not exceed 12 degrees in both cases, which is a good result, considering that the neural network was trained with an azimuth step of 10 degrees. Mostly the same accuracy of localization in the new room, and room where the train dataset was recorded, indicates a good generalization property for the trained neural network.

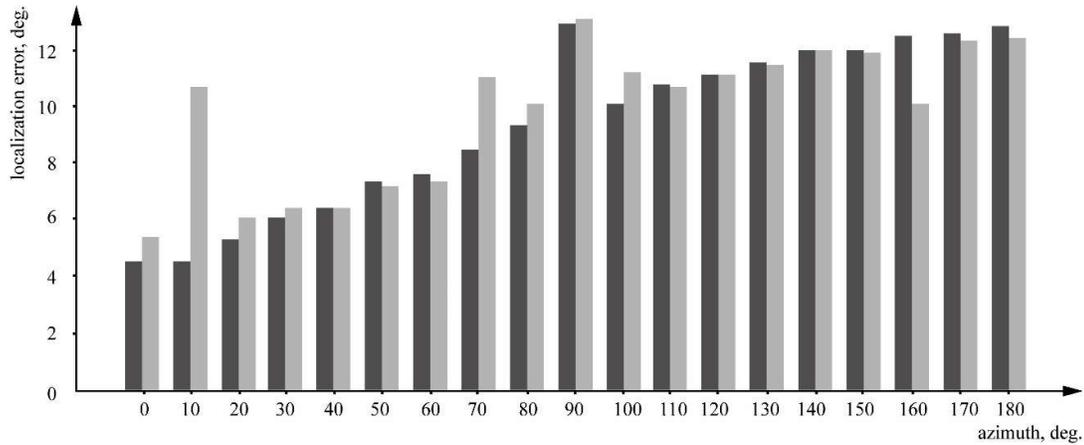

Fig. 6 Dependence of average absolute values of azimuth determination error from real azimuth with active sound source. Light grey bars show measurements conducted in the new room. Dark grey bars show measurements conducted in the room where the training dataset was recorded.

Fig. 7 gives an example of results of continuous localization of a stationary source that plays music. It can be seen that localization error is complexity of choice between neighbouring classes of neural networks.

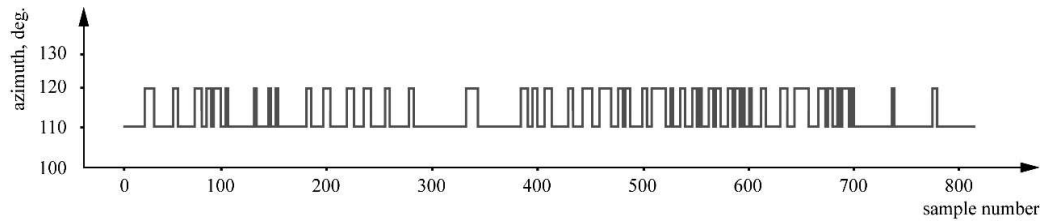

Fig. 7 Values of calculated azimuths for an active sound source as a function of sound frame number with real direction of 110 degrees.

**Impact on the whole speech recognition pipeline:** A comparison of the accuracy of far-field speech recognition was performed with three configurations of speech recognition pipeline to understand the impact of developed sound source localization on the final result of speech recognition.

The first speech recognition pipeline didn't use microphone array processing:

1. Audio capturing from the first channel of the microphone array.
2. Voice activity detection using code from the WebRTC project.
3. Speech recognition using Google Speech API.

The second speech recognition used sound source localization developed by [3]:

1. Audio capturing from the microphone array.
2. Sound source localization using weighted GCC-PHAT with Kalman filtering.
3. MVDR beamformer [7].
4. Zelinski post-filter [21].
5. Speech recognition using Google Speech API.

Implementations of MVDR beamformer and Zelinski post-filter were used from BTK toolkit. And the last speech recognition pipeline used developed sound source localization based on the residual network:

1. Audio capturing from the microphone array.
2. Proposed sound source localization using residual network with Kalman filtering.
3. MVDR beamformer.
4. Zelinski post-filter.
5. Speech recognition using Google Speech API.

100 phrases were recognized simultaneously through 3 described speech recognition pipelines. Speech recognition pipelines shared the same microphone array during the experiment. Voice sound sources were located on distance 1.5 m from the microphone array at different directions. Word error rates (WER) were calculated and compared for results from pipelines (Table 1).

Table 1 WER value for different configurations of speech recognition pipeline

| № | Speech recognition pipeline | WER, % |
|---|---|---|
| 1 | Mono audio capturing without any speech enhancement. | 2.21 |
| 2 | Speech enhancement using beamforming and GCC-PHAT sound source localization. | 2.99 |
| 3 | Speech enhancement using beamforming and proposed sound source localization. | 1.85 |

The best result was shown by the solution with proposed sound source localization. High WER is shown by the solution with GCC-PHAT because the width of the beam pattern formed by the MVDR beamformer is lower than the accuracy of the sound source localization achieved GCC-PHAT on used microphone array, so sometimes beam pattern became orientated not to the sound source. The width of the beam pattern of the MVDR beamformer is about 20° (Fig. 8). Average localization error of the developed sound source localization system is not higher than 12°. So, their combination achieves a good quality of speech enhancement resulting in low WER. Directivity pattern was modeled using following equations [22]:

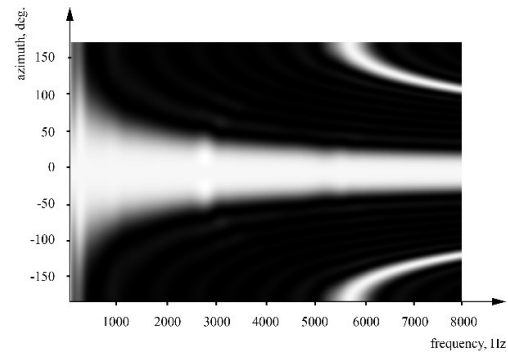

Fig. 8 The beam pattern of the MVDR beamformer in the endfire orientation for used microphone array.

$$\Psi(\omega, \Theta) = |H(\omega)^H u(\omega, \Theta)|^2 \qquad (22)$$

MVDR filter coefficient vector [22]:

$$H(\omega) = \frac{\Phi(\omega)_{NN}^{-1} a(\omega)}{a(\omega)^H \Phi(\omega)_{NN}^{-1} a(\omega)} \quad (23)$$

The noise cross-power spectral matrix [7]:

$$\Phi_{NN}(\omega) = \Phi_{N'N'}(\omega) + \Phi_{II}(\omega) \quad (24)$$

The instrumental noise cross-power spectral matrix [6]:

$$\Phi_{II}(\omega) = N_I^2(\omega) I \quad (25)$$

, where $N_I(\omega)$ is the magnitude of the instrumental noise in single microphone.

The cross-spectral density for an isotropic noise field [7]:

$$\Phi_{ij}(\omega) = N_O(\omega) sinc(\frac{\omega d_{ij}}{v}) \quad (26)$$

$$\Phi_{N'N'}(\omega) = \begin{bmatrix} \Phi_{11}(\omega) & \cdots & \Phi_{1M}(\omega) \\ \vdots & \ddots & \vdots \\ \Phi_{M1}(\omega) & \cdots & \Phi_{MM}(\omega) \end{bmatrix} \quad (27)$$

, where $N_O(\omega)$ is the noise spectrum captured by an omnidirectional microphone, v is the speed of sound, $d_{ij}$ is a distance between i and j microphones.

The propagation vector for linear microphone array [22]:

$$a(\omega) = \left[ a_i : e^{\frac{j\omega p_i}{v}}, i = 1, \ldots, M \right] \quad (28)$$

, where $p_i$ is a position of i microphone.

The unit vector in the required direction of beam pattern [22]:

$$u(\omega, \Theta) = \left[ u_i : e^{\frac{j\omega \cos(\Theta) p_i}{v}}, i = 0, \ldots, M-1 \right] \quad (29)$$

## 6 Conclusion

A sound source localization method based on deep residual neural networks was developed. It doesn't require a captured signal to be transformed from time domain to frequency domain with Fourier transformation, which positively affects system performance. The developed method demonstrated good accuracy of the sound source direction azimuth determination with a linear compact microphone array even without the consideration of object dynamics with a Kalman Filter or Particle Filter. As a further improvement to the method, the system can be trained in such a way that will allow us to determine several sound source locations simultaneously. Also in future work, the architecture should be complemented by LSTM, BLSTM or GRU layers [23], to make the network able to consider object dynamics.

## ACKNOWLEDGMENTS

This work was supported by the Russian Innovation Support Fund (project 102GRNTIS5/26071).

## REFERENCES

1. Woelfel, M., McDonough, J., 2009. Distant Speech Recognition. John Wiley & Sons, The City of New York.
2. Kumatani, K., McDonough, J., Raj, B., 2012. Microphone array processing for distant speech recognition: From close-talking microphones to far-field sensors. IEEE Signal Processing Magazine, 29(6):127-140. http://dx.doi.org/10.1109/MSP.2012.2205285
3. Grondin, F., Michaud, F., 2015. Time difference of arrival estimation based

on binary frequency mask for sound source localization on mobile robots. IEEE/RSJ International Conference on Intelligent Robots and Systems, Hamburg, Germany. http://dx.doi.org/10.1109/IROS.2015.7354253
4. Tashev, I., Acero, A., 2006. Microphone array post-processor using instantaneous direction of arrival. International Workshop on Acoustic, Echo and Noise Control IWAENC, Paris, France.
5. Valin, J.M., Michaud, F., Rouat, J., 2007. Robust localization and tracking of simultaneous moving sound sources using beamforming and particle filtering. Robotics and Autonomous Systems Journal (Elsevier), 55(3):216- 228.
6. Ishi, C., Chatot, O., Ishiguro, H., et al., 2009. Evaluation of a MUSIC-based real-time sound localization of multiple sound sources in real noisy environments. IEEE/RSJ International Conference on Intelligent Robots and Systems, St. Louis, USA, p.2027-2032. http://dx.doi.org/10.1109/IROS.2009.5354309
7. Tashev, I., 2009. Sound Capture and Processing: Practical Approaches. John Wiley & Sons, The City of New York.
8. Yalta, N., Nakadai, K., Ogata, T., 2017. Sound source localization using deep learning models. Journal of Robotics and Mechatronics, 29(1):37-48. http://dx.doi.org/10.20965/jrm.2017.p0037
9. Suvorov, D., Zhukov, R., Evmenenko, A., et al., 2017. Device for audiovisual temperature-invariant voice source localization (in Russian). Russian Patent, 170249.
10. Ronzhin, A., Karpov, A., 2008. Comparison of methods for localization of multimodal system user by his speech (in Russian). Journal of Instrument Engineering, 51(11):41-47.
11. Suvorov, D., Zhukov, R., 2017. Device for synchronous data capturing from the array of MEMS microphones with PDM interface (in Russian). Russian Patent, 172596.
12. Tzanetakis, G., Cook, P., 2002. Musical genre classification of audio signals. IEEE Transactions on Audio and Speech Processing, 10(5):293-302. http://dx.doi.org/10.1109/TSA.2002.800560
13. Eren, L., 2017. Bearing fault detection by one-dimensional convolutional neural networks. Mathematical Problems in Engineering, 2017.

http://dx.doi.org/10.1155/2017/8617315

14. He, K., Zhang, X., Ren, S., et al., 2016. Deep residual learning for image recognition. IEEE Conference on Computer Vision and Pattern Recognition (CVPR), Las Vegas, USA, p.770-778. http://dx.doi.org/10.1109/CVPR.2016.90

15. Ioffe, S., Szegedy, C., 2015. Batch normalization: Accelerating deep network training by reducing internal covariate shift. 32nd International Conference on Machine Learning (ICML-15), Lille, France, p.448-456.

16. Srivastava, N., Hinton, G., Krizhevsky, A., et al., 2014. Dropout: A simple way to prevent neural networks from overfitting. Journal of Machine Learning Research, 15(1):1929-1958.

17. Maas, A., Hannun, A., Ng, A., 2013. Rectifier nonlinearities improve neural network acoustic models. ICML Workshop on Deep Learning for Audio, Speech and Language Processing, Atlanta, USA, 28.

18. McGregor, S., 2007. Neural network processing for multiset data. 17th international conference on Artificial neural networks ICANN 2007, Porto, Portugal, p.460-470.

19. Kingma, D., Ba, J., 2015. Adam: a method for stochastic optimization. The International Conference on Learning Representations ICLR, San Diego, USA.

20. Maaten, L., Hinton, G., 2008. Visualizing data using t-SNE. Journal of Machine Learning Research, 9(1):2579-2605.

21. Aleinik, S., 2017. Acceleration of Zelinski post-filtering calculation. Journal of Signal Processing Systems, 88:463- 468. http://dx.doi.org/10.1007/s11265-016-1191-9

22. Vary, P., Martin, R., 2006. Digital Speech Transmission: Enhancement, Coding and Error Concealment. John Wiley & Sons, The City of New York. http://dx.doi.org/10.1002/0470031743

23. Chung, J., Gulcehre, C., Cho, K., et al., 2014. Empirical evaluation of gated recurrent neural networks on sequence modeling. NIPS 2014 Workshop on Deep Learning, Montreal, Canada.